\documentclass[12pt,preprint]{aastex}

\shortauthors{Di Stefano \& Kilic} 
\shorttitle{Ex-companions in SNe Ia Remnants}
\def\susd{spin-up/spin-down}  \def\ia{SN~Ia} \def\iae{SNe~Ia}

\begin{document} \title{The Absence of Ex-Companions in Type Ia Supernova Remnants}

\author{R. Di Stefano\altaffilmark{1} and Mukremin Kilic\altaffilmark{2}}

\altaffiltext{1}{Harvard-Smithsonian Center for Astrophysics, 60 Garden St.,
Cambridge, MA 02138, USA; rd@cfa.harvard.edu}

\altaffiltext{2}{Homer L. Dodge Department of Physics and Astronomy, University
of Oklahoma, 440 W. Brooks St., Norman, OK, 73019, USA; kilic@ou.edu}

\begin{abstract}

Type Ia supernovae (\iae) play important roles in our study of the expansion and
acceleration of the Universe, but because we do not know the exact nature or
natures of the progenitors, there is a systematic uncertainty that must be
resolved if SNe Ia are to become more precise cosmic probes. No progenitor
system has ever been identified either in the pre- or post-explosion images of a
Ia event. There have been recent claims for and against the detection of
ex-companion stars in several \iae\ remnants. These studies, however, usually
ignore the angular momentum gain of the progenitor WD, which leads to a spin-up
phase and a subsequent spin-down phase before explosion. For spin-down
timescales greater than $10^5$ years, the donor star could be too dim to detect
by the time of explosion.  Here we revisit the current limits on ex-companion
stars to SNR 0509-67.5, a 400 year old remnant in the Large Magellanic Cloud
(LMC). If the effects of possible angular momentum gain on the WD are included,
a wide range of single-degenerate progenitor models are allowed for this
remnant.  We demonstrate that the current absence of evidence for ex-companion
stars in this remnant, as well as other \iae\ remnants, does not necessarily
provide the evidence of absence for ex-companions. We discuss potential ways to
identify such ex-companion stars through deep imaging observations.

\end{abstract}

\keywords{accretion, accretion disks -- supernovae:general -- supernova remnants
-- white dwarfs}

\section{Introduction}

No progenitor system for a Type Ia supernova (SN Ia) has ever been identified,
nor do we know the exact nature or natures of the progenitors.  We do know that
SNe Ia are explosions of carbon/oxygen core white dwarfs (WDs), and that is
the extent of our current knowledge on the progenitor systems.
We do not know the type of astrophysical system within which the WD comes to gain mass.
One possibility is a binary in which the
WD accretes mass from a non-degenerate companion star in the single-degenerate
(SD) channel \citep{whelan73}; another is through the merger of two WDs in the
double-degenerate (DD) channel \citep{iben84,webbink84}. 
Yet another is the ``double-detonation'' model, in which a sub-Chandrasekhar mass WD
detonates following a detonation of an accumulated helium layer on the surface
\citep[][and references therein]{nomoto80,woosley94,ruiter11,sim12}. Resolving this issue is
important because the characteristics of the companion and the mass transfer
process create the environment within which the explosion occurs.

Theoretical arguments can be marshalled to support or reject the hypothesis that
most \iae\ are produced through the SD channel, and the same is true for the DD
channel. For example, SDs can produce explosions with the characteristic typical
of \iae\ \citep[e.g.][]{hillebrandt00,livio00}, while it has proved difficult to
do the same for DD models. Even though the majority of the DD mergers may result
in an accretion induced collapse to a neutron star \citep{nomoto91}, some of
these mergers may instead produce \iae\ \citep{yoon07,pakmor10,dan11}.  While
neither channel appears capable of reproducing the measured rates of \iae\ in
galaxies, the distribution of times after star formation associated with the DD
models is a better fit to the measured delay time distribution \citep{maoz10}.

Observational evidence has not yet been able to provide a resolution.  The mass
distribution of WDs in the solar neighborhood provides some evidence that
mergers occur \citep{liebert05}. Yet, while radial velocity surveys have
identified a significant binary WD merger population \citep{kilic10,brown12},
there is no confirmed binary WD system with a total mass above $M_{Ch}$ ($\approx 1.4 M_{\odot}$) and a
merger time shorter than a Hubble time.  Observational tests of the SD scenario
would seem potentially more direct.  Several post-explosion investigations, for
example, detect circumstellar material likely to have been ejected from the
progenitor binary \citep[e.g.][]{patat07,sternberg11}.  Observations of the
supernova remnant, RCW 86, which appears to have been associated with an \ia ,
find features best explained by an explosion into a cavity created by a
single-degenerate progenitor system.  However, the true nature of RCW 86,
whether it is a core-collapse or a Ia explosion, remains questionable \citep[see
the discussion in][]{williams11}.

SNe~Ia are rare, so we discover them at large distances from us.  Furthermore,
the progenitor systems, even in SD models, are predicted to be dimmer than
typical progenitors of core-collapse supernovae, therefore impossible to detect
in pre-exlosion images of most SNe~Ia host galaxies.  Thus, nearby \iae\ are
particularly valuable to progenitor studies.  The supernova 2011fe occurred in
the galaxy M101, which, at $6.4$~Mpc, is relatively nearby. Pre-explosion images
were able to rule out the presence of a giant donor star
\citep{nugent11,li11,bloom12,horesh12}; the results of radio observations are
consistent with this conclusion, placing limits on pre-explosion winds
\citep{chomiuk12}.  All other SD models, including those with subgiant or
main-sequence donors, are allowed by the full complement of pre- and
post-explosion observations.

One of the last SNe~Ia to occur in the Milky Way was ``Tycho's'' supernova (SN
1572), whose remnant is measured to be $2.5-3$~kpc away.  This is close enough
to allow us to discover the widowed companion to the exploded WD, should it
exist. \citet{ruiz04} identified a G2 dwarf as the ex-companion. However, this
identification remains controversial \citep{kerzendorf09,gonzalez09}.

Recently, \citet{schaefer12} extended the progenitor search to a supernova
remnant in the LMC, SNR~0509-67.5. Based on a relatively short {\em Hubble Space
Telescope (HST) Wide Field Camera 3} exposure of the SNR, they rule out any
ex-companion star brighter than V = 26.9 mag in the central error circle. Given
the distance to the LMC, this corresponds to sources brighter than $M_V=+8.4$
mag.  \citet{schaefer12} argue that this magnitude limit excludes all SD
progenitors and that the only remaining possibility is that of a DD merger
system, with no companion left behind.

In this paper we point out that, if the effects of angular momentum carried by
matter falling toward the WD are included, a wide range of SD progenitor models
are allowed. These \susd\ models are reviewed in \S 2. In \S 3 we revisit the
analysis of possible widowed donors in SNR~0509-67.5, and review the
implications for the study of other \iae\ remnants in \S 4.

\section{Spin-up/Spin-down Models}

\subsection{Angular Momentum} 

The term ``single-degenerate model'' refers to
diverse classes of binary systems within which a WD may accrete matter and
achieve the critical mass.  The donor may be drawn from a broad range of stellar
types. Orbital separations may be close enough that the donor fills its Roche
lobe, or wide enough that mass transfer occurs through winds. Matter may be
accreted through a disk, or not.  The common feature is that the WD should be
able to retain enough of the matter it accretes to allow it to eventually
achieve the critical mass needed for explosion.

Mass retention is generally thought to require that incoming matter undergo
nuclear burning. Quasi-steady nuclear burning requires high rates of accretion,
generally above a few times $10^{-7} M_\odot$~yr$^{-1}$ \citep{iben82,nomoto82}.
Matter can also be retained with the somewhat lower accretion rates consistent
with recurrent novae \citep{prialnik95}. Note however, that high rates of mass
infall would be required even if nuclear burning were not an issue, because the
types of donors which can donate enough mass to a CO WD to allow it to achieve
$M_{crit}$ would necessarily donate mass at high rates for extended periods
\citep{rappaport94}.

In almost all accretion scenarios, infalling matter carries angular momentum.
The WD therefore accretes both mass and angular momentum. While the accretion of
mass has been considered in detail, in terms of both the burning of infalling
mass and the evolution toward explosion, the increase in angular momentum is a
difficult problem that has been less well studied.  While results for some
detailed models have been computed \citep[e.g.,][]{yoon04}, there are
uncertainties about important physical elements, such as the relative roles of
viscosity, magnetic fields, and gravitational radiation in promoting both
spin-up and spin-down \citep[see e.g.,][]{piro08}.  Recently, the signficance of
the issue of angular momentum transfer has been linked to the appearance of the
progenitor binary both pre- and post-explosion, using physical arguments
independent of the details of the relevant processes \citep{distefano11,justham11}.

It is important to note that much theoretical and observational work needs to be
done to understand the effects of angular momentum. For example, several sets
of calculations \citep[e.g.,][]{iben82,nomoto82,shen07,prialnik95} have been
conducted to determine the range of accretion rates that allow the WD to burn
incoming material or that are associated with nova. These have all assumed spherical
symmetry, which does not hold if the WD spins at a rate of a few tenths of a
Hertz. Furthermore, the calculations apply to WDs mith mass below the
Chandrasekhar mass. Thus, the accretion calculations must be extended if we are to
understand the details of how matter accretes onto spinning WDs of all masses,
including massive WDs. Several other questions must also be addressed, including
the effects of a delay time to explosion on the internal state of the WD. On the
observational front, searches for fast spinning WDs are needed. In this paper we
focus on the specific issue of the effects spin-up/spin-down would have on the
detectability of the widowed companion to a WD that explodes.

\subsection{Spin-Up/Spin-Down Models}

\citet{distefano11} introduced classes of spin-up/spin-down binary models for
accreting WDs, and studied their implications for the detectability of the
widowed companions. Briefly, in the SD channel, SNe~Ia explosions result from
WDs retaining the accreted mass. The angular momentum of the WD must increase
because of accretion.  There are several examples of fast spinning WDs in close
binary systems, mostly in intermediate polars, e.g., the dwarf nova WZ Sge
(27.87 s), the nova-like AE Aqr (33.06 s), and the nova remnant V842 Cen (56.82
s). The likely progenitors of SNe Ia have mass transfer rates that are several
orders of magnitude higher than those inferred for intermediate polars. The
retention of mass should spin mass-gaining WDs to even shorter periods than
measured for intermediate polars. The WD in HD 49798 is one such system; it has
a 13-s spin period 1.28 $M_{\odot}$ WD with a hot subdwarf companion
\citep{mereghetti11}.

Spin-up can increase the value of the critical mass, $M_{crit}$  needed for
explosion \citep{ostriker66,anand65,roxburgh65}.  If $M_{crit}$ has increased,
the WD may not explode even after its mass has become equal to $M_{Ch}$.  If the
mass of the WD cannot reach $M_{crit}$, then the explosion must be delayed until
the WD can spin down, reducing the value of $M_{crit}$ to the actual value of
the WD's mass.  By ``spin-down'' we mean that the WD has either lost angular
momentum or changed its internal angular momentum profile so that the required
central density can be achieved. Spin-down begins when the mass transfer rate
becomes low enough that genuine mass gain is no longer possible.

The need for a spin-down time, can mean that the donor has different properties
at the time of explosion from those it would have had had the explosion occurred
at the time when the mass  of the WD reached $M_{Ch}.$ 

\subsubsection{Giant Donors}

Consider the case in
which the donor is a giant during the phase in which it contributes mass at a
high enough rate to promote nuclear burning. Because a giant loses mass at
higher rates as it evolves, the WD is very likely to continue gaining mass and
angular momentum until the envelope of the giant donor is exhausted. At that
point, mass transfer ceases, and the first-formed WD can begin to spin down and
to cool after mass transfer, just as its companion, the degenerate core of the
donor star, also begins to cool. At the time of explosion, the remnant of the
donor is a WD, less massive than the one that exploded. It has been cooling for
a time equal to the time the massive WD needed to spin down.  Its mass and
temperature can therefore be used to measure the spin-down time. \citet{justham11}
argues that the amount of hydrogen left on the donors in these systems is below the
current detection limits, providing a potential explanation for the lack of hydrogen
in \iae\ spectra.

\subsubsection{Main-sequence Star Donors}

Consider the case in which the donor is a main-sequence star. In this case the
donor must be more massive than the WD when it first fills its Roche lobe, in
order for enough mass to be donated at an accretion rate high enough to promote
nuclear burning and to bring a carbon-oxygen WD to the Chandrasekhar mass. As
the mass of the donor star decreases and the mass of the WD increases, the rate
of mass transfer declines. Once the mass ratio reverses, mass transfer continues
at a high rate until the donor star achieves thermal equilibrium. During this
interval, much of the accreted mass continues to be burned; instead of being
processed in a quasisteady manner, however, it may burn episodically during
recurrent novae.  Once the donor star achieves thermal equiliburium, its thermal
readjustment no longer drives mass transfer, which must then proceed through the
agency of magnetic braking, producing a mass-transfer rate that starts at about
$10^{-8} M_\odot$~yr$^{-1}$ and declines as the donor continues to lose mass.
The WD can no longer retain the accreted mass and is free to spin down.  By the
time of explosion, the companion star may still be donating mass at a low rate.
Just prior to explosion, the system will be a cataclysmic variable (CV), with a
low-mass donor. The mass of the donor can provide a rough measure of the
spin-down time.

\subsubsection{Subgiant Donors}

If the donor is a subgiant at the time during which it donates mass at a
high-enough rate to promote lasting mass gain by its WD companion, then elements
of both scenarios apply. Like the case in which the donor is a main-sequence
star, the mass accretion rate will decline sometime after the mass ratio
reverses. But, as in the giant-donor case, the donor will run out of envelope
after an interval of mass transfer.  Thus, depending on the spin-down time, the
system may be a catclysmic variable (similar to AE Aqr, which we will discuss
below) just prior to explosion, or else it may consist of a WD (likely a He WD)
orbiting the massive WD.

\subsubsection{Spin-down Time}

The general trend is that the donor star will be less massive and less luminous
at the time of explosion if spin-up/spin-down occurs. The physical
characteristics of the system are determined by the spin-down time.  The
relationship between the properties of the widowed donor to the spin-down time
is fortunate, because it has the potential to provide a way to estimate the
spin-down time of real systems. First-principles estimates have proved
difficult, and the spin-down time is not well constrained at present.
Theoretically, spin-down times are uncertain by several orders of magnitude;
estimates range from $\sim$$10^5$ to $>$$10^9$ years \citep{lindblom99,yoon05}.
For example, \citet{vanamerongen87} show that the timescale for the secular
decrease of the spin rate of an accreting magnetized WD in CVs is ($6 \times
10^8 yr)(P_{orb}/4 hr)^{-2.2}$.

Unfortunately, observations of spinning-down WDs are scarce. The best example of
a WD currently spinning down is AE Aqr, which contains a 0.79 $\pm$ 0.16
$M_{\odot}$ WD and a 0.50 $\pm$ 0.10 $M_{\odot}$ K5 dwarf companion in a 9.9 h
orbit \citep{casares96}.  The WD in AE Aqr is currently spinning down at a rate
$\dot P = 5.64 \times 10^{-14}$ s s$^{-1}$ \citep{dejager94}, which corresponds
to a spin-down time of $2 \times 10^7$ years \citep{mauche06}.  The evolutionary
history of AE Aqr exactly parallels that of SD progenitors that experience spin
up: it apparently had a high mass transfer rate, allowing the WD to gain mass;
it is now spinning down. The mass gained was, however, not enough to bring the
WD to a mass near $M_{Ch}$.  There is evidence that other WDs in CVs have gained
mass.  For example, \citet{yuasa10} study the mass distribution of WDs in
intermediate polars and show that the average WD mass of their sample is 0.88
$\pm$ 0.25 $M_{\odot}$.  This is significantly larger than the average mass of
the WDs in the solar neighborhood \citep{tremblay11}. In addition, seven of the
17 systems in the \citet{yuasa10} study have WD masses above 1 $M_{\odot}$.  AE
Aqr is perhaps the best example of a CV with a WD of mass smaller than $M_{Ch}$
that appears to have followed the evolutionary track we suggest.  In fact, since
binaries in which a WD achieves the critical mass must be a small minority of
binaries in which the WD accretes matter, we expect AE Aqr to be an example of
many other members of its class. This system demonstrates that, even though
angular momentum and its attendant physics are difficult to compute,
spin-up/spin-down is realized in nature.

\section{SNR 0509-67.5}

SNR 0509-67.5 is a 400-year-old SN Ia remnant in the LMC. Based on its light echo
and X-ray spectrum, \citet{rest08} and \citet{badenes08} demonstrate that
SNR 0509-67.5 originated from an exceptionally bright \ia\ that synthesized
$\sim 1 M_{\odot}$ of $^{56}$Ni. Its youth and
location means that it provides a unique opportunity to determine the nature of
its progenitor and to take an important step toward resolving the so-called
progenitor puzzle. It is possible the WD whose explosion created SNR 0509-67.5
accreted mass from a companion. If so, that donor star likely survived the
explosion, and is moving away from its site with a speed comparable to its
former orbital speed. It would presently lie within about $0.3\arcsec$ of the
explosion site, if its transverse velocity is smaller than $200$~km~s$^{-1}$.
The LMC is close enough that we can, in principle, detect this newly ``widowed''
donor star.

Using 1010, 696, and 800 s exposures in the $B, V$, and $I$ filters taken with
the {\em HST Wide Field Camera 3}, \citet{schaefer12} rule out ex-companion
stars down to $V=26.9$ mag in the central 1.4$\arcsec$ region.  This magnitude
limit corresponds to $M_V=+8.4$ mag at the LMC's distance, ruling out a wide
range of possible donors.  This limit, however, does not take into account the
extinction in the line of sight to the LMC ($A_V\approx0.25$ mag, A. Pagnotta
2012, private communication). Hence, the absolute magnitude limit achieved by
the current {\em HST} observations is at $M_V=8.15$ mag, as shown in Figure 1.

\begin{figure} 
\plotone{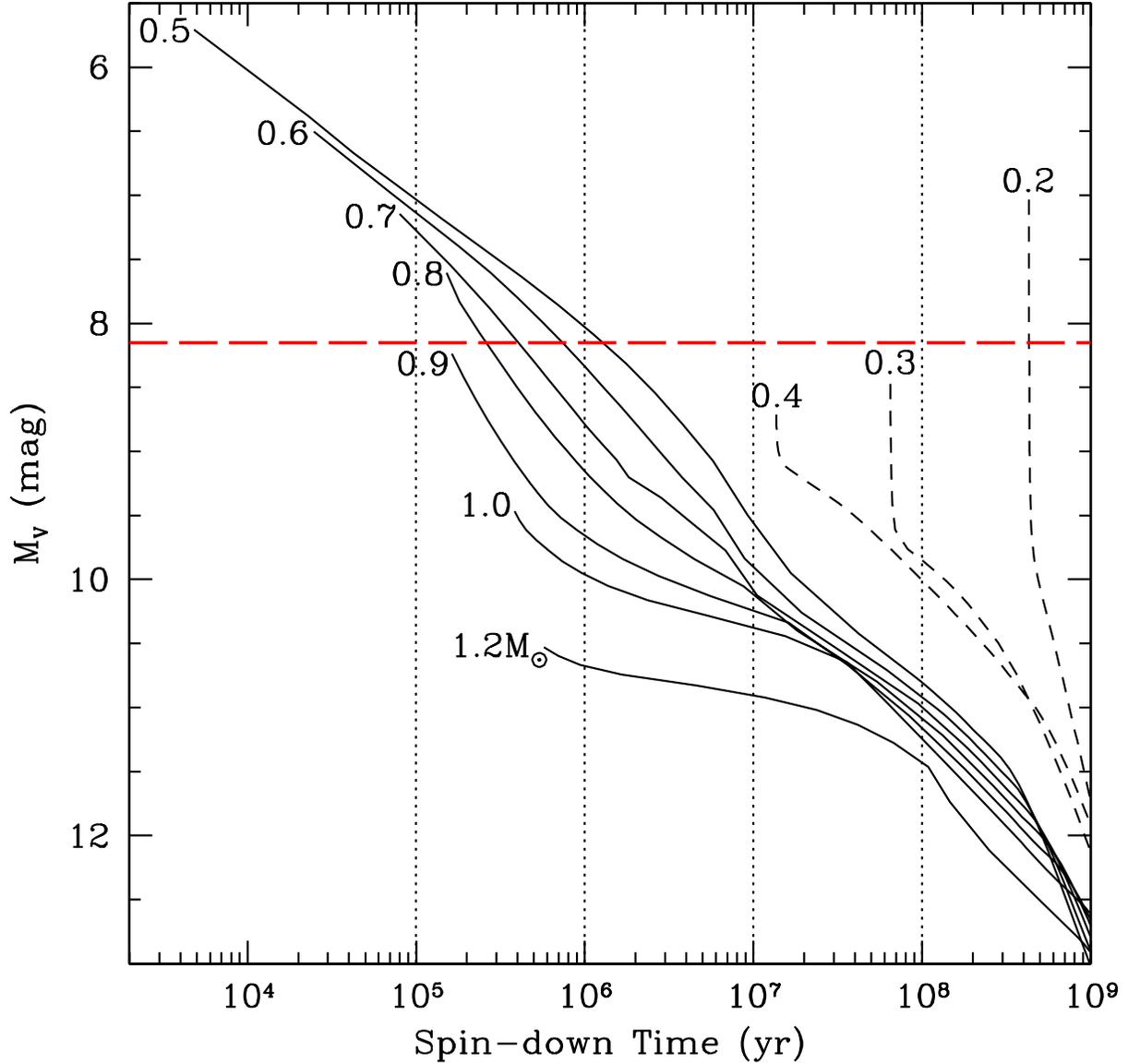}
\caption{The absolute magnitude versus the (cooling age) spin-down time for
ex-companion WDs.  The solid and dashed lines show the cooling curves for a
variety of C/O core \citep{bergeron11} and low-mass He-core WDs
\citep{serenelli01}.  The He-core models include the effects of element diffusion
and hydrogen shell flashes, which lead to thinner hydrogen envelopes and cause the evolutionary
path shown \citep{althaus01}. \citet{schaefer12} obtain an absolute $V-$band magnitude
limit of +8.4.  However, this limit does not take into account the extinction in
the line of sight to the LMC ($A_V\approx0.25$ mag, A. Pagnotta 2012, private
communication). Hence, the correct absolute magnitude limit achieved by the
current {\em HST} observations is at $M_V=8.15$ mag, which is shown by the long
dashed line.} \end{figure}

To use this limit to place constraints on the flux from the widowed donor, it is
important to locate the position at which the explosion occurred. We need to be
certain that no stars lie close enough to the explosion site to be identified as
possible ex-donors.  Had the event occurred in a uniform medium, the supernova
remnant would likely exhibit axial symmetry, making it possible to uniquely
identify a center, which would be the most likely point of explosion. In the
case of SNR 0509-67.5, it exhibits an asymmetric spatial pattern indicating that
the surrounding medium is not uniform. Simulations can reproduce the pattern and
identify the true site of the explosion (S. Reynolds 2012, private
communication).  \citet{schaefer12} instead studied the geometry of the remnant
to identify the likely explosion site, in a manner that also considers the
asymmetry. Should the supernova have occurred at a position offset from the one
identified geometrically, then, instead of a definite null detection, there
could be a set of candidates for the widowed donor, shown in Figure 1 of
\citet{schaefer12}.  In this case, each candidate would correspond to a possible
binary model, and each would place different possible limits on the spin-down
time. If one of these candidates was observed to have high proper motion and/or
to be enriched in materials expected from the explosion, then the progenitor
would be identified and the binary model could be unique. In the discussion below,
we assume that the geometrical work produces the correct result, so that there are
no candidate ex-donors down to the flux probed by the HST observations. If this is
not the case, and there are candidate donors, then each would need to be studied
separately. This more complex process will certainly apply to the sites of other
remnants of SNe~Ia.

\citet{schaefer12} conclude that all possible donor stars had been eliminated,
proving that the progenitor was a DD system whose components merged\footnote{However,
helium-rich degenerate donor stars in the double-detonation model
\citep[e.g.,][]{nomoto80,woosley94,ruiter11,sim12} cannot be ruled out.}.  
If, however, the WD gained mass from a companion, it would likely have also gained
angular momentum. It may therefore have a higher critical mass and may need to
spin down to decrease the value of $M_{crit}$ to the current value of its mass.

After the spin-down episode, the donor star could be dim enough by the time of
explosion that it would not have have been detected in the existing data.  The
unknown spin-down timescales for fast spinning massive WDs are therefore
extremely important for the visibility of ex-companion star in SNR 0509-67.5.
Here we consider two cases, a red-/sub-giant companion and a main-sequence star
companion.

\subsection{Red Giant Companions}

For giants and some subgiants, the mass transfer rate will not decrease, and
spin-down cannot start before the giant's envelope is depleted. The giant donor
becomes a WD that cools during spin down \citep{distefano11,justham11}. Figure 1 displays
the absolute magnitude versus the (cooling age) spin-down time for ex-companion WDs. 
If the spin-down takes longer than $10^5$ years, such ex-donors are likely to be
very dim today, requiring deep imaging observations. Ruling out present-day He-core
($M < 0.45 M_{\odot}$) WDs would rule out an ex-donor on the red giant branch;
allowing higher mass present-day WDs would indicate ex-donors on the asymptotic giant branch.

For spin-down times of $10^6$ years, the current {\em HST} observations allow
ex-donors that are WDs with $M>0.6 M_{\odot}$. For spin-down times of $10^7$
years, only low-mass He-core WDs are ruled out; higher mass WDs with
$M>0.4M_{\odot}$ would be hidden in the current data. For even longer spin-down times of
$10^8-10^9$ years, the current data cannot rule out even the lowest mass He-core
WDs. Figure 2 displays the ruled-out ex-companion WDs for a given spin-down
time; C/O core WDs may be hiding in the center of this remnant if the spin-down
times are longer than about $10^5$ years.

\begin{figure} 
\plotone{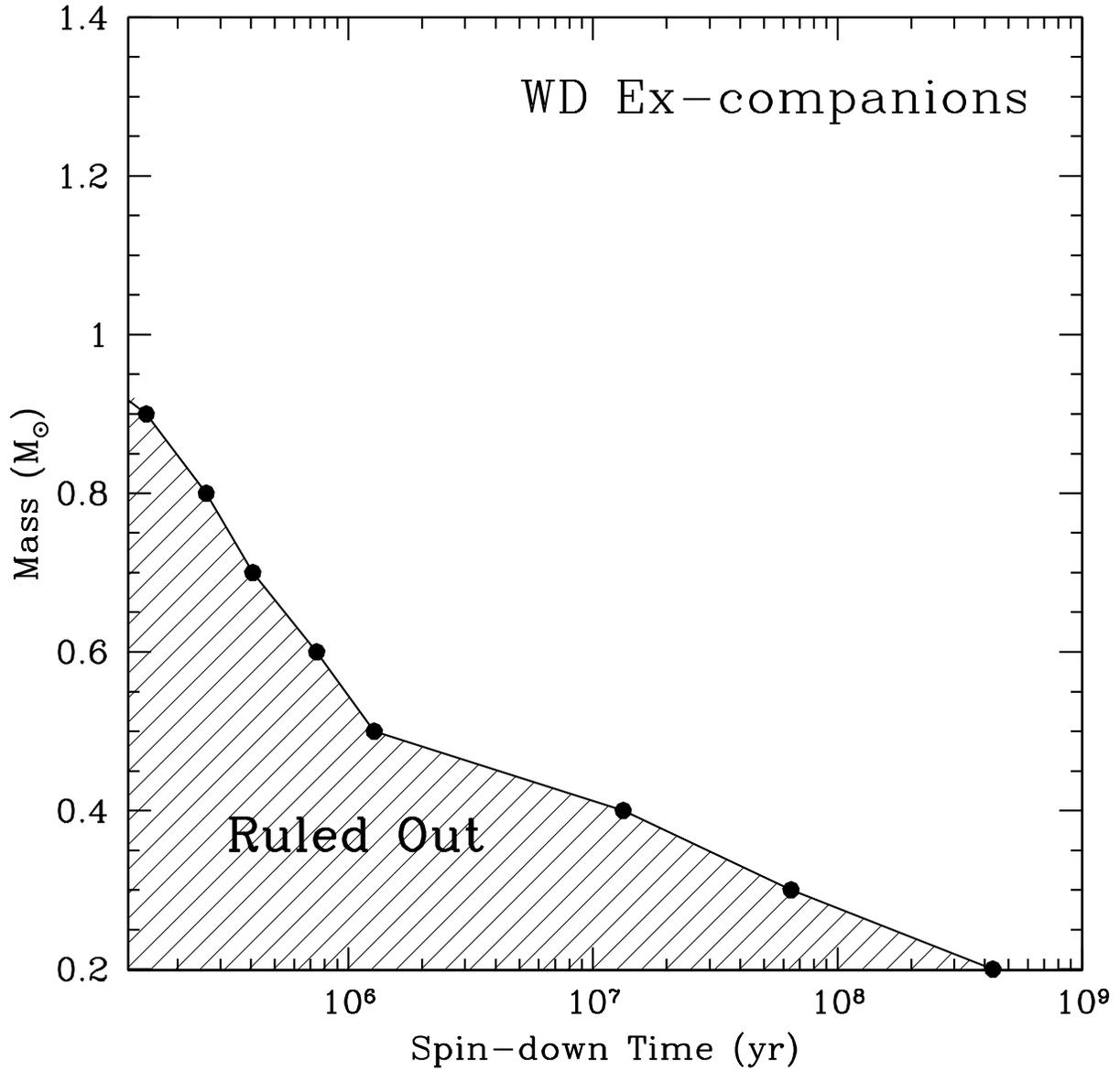}
\caption{Masses of the potential ex-companion WDs to SNR 0509-67.5. The shaded
area marks the ruled-out companions based on the currently available {\em HST}
imaging data.} \end{figure}

\subsection{Main-sequence Companions} 

Figure 3 shows the evolution of the
ex-companion main-sequence stars to (super-)Chandrasekhar-mass
WD explosions.  We consider 1.79 and 1.39 $M_{\odot}$ WDs. For each
value of the WD mass, we have assumed that the low-$\dot M$ phase begins when
the donor star is $0.1\, M_\odot$ less massive than its WD companion.  This
value is uncertain, since it depends on issues such as the readjustment of the
donor star, the generation of radiation from the vicinity of the accreting WD,
and its effect on the donor. Once magnetic braking becomes the dominant driver
of orbital evolution and mass transfer, the system will have entered the realm
of CVs, which have been well studied, using both theory and observations.
Nevertheless, there are significant uncertainties about key issues such as
magnetic braking, and the possible ``bloating'' of the donor star.  To
illustrate that there is a range of possibilities, we show, for each value of
the initial donor mass, the results for two different values of the magnetic
braking parameter, $\gamma$ \citep[cf. Eq. 36 in][]{rappaport83}, as described in the caption.

\begin{figure} 
\plotone{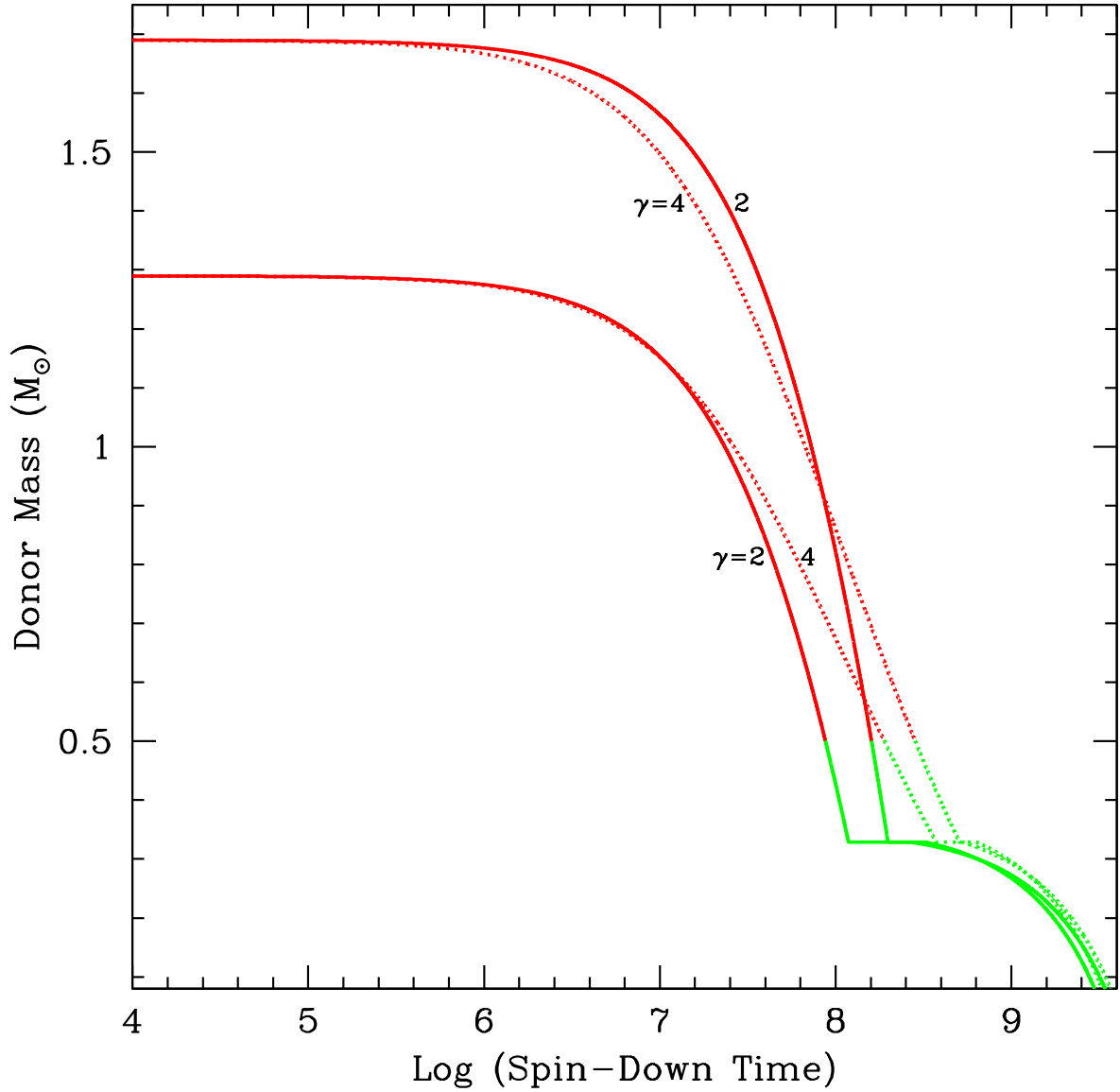}
\caption{Masses of the potential ex-companion main-sequence stars to 1.8 $M_{\odot}$
and 1.4 $M_{\odot}$ WDs for $\gamma=2$ (solid lines) and $\gamma=4$
(dotted lines), respectively.  The current {\em HST} imaging data
on SNR 0509-67.5 rule out $\gtrsim0.5 M_{\odot}$ ex-companions (red curves).}
\end{figure}

The limits placed by a non-detection in the current data mean that a
main-sequence star with mass larger than approximately $0.5\, M_\odot$ would
have been detected if the spin-down time is shorter than about $10^{7.9}$~years
to $10^{8.3}$ years. Thus, it is possible that the donor star would not have
been detected, if the spin-down time is in this range or longer.

It is important to push the limit on the mass of the widowed donor to even lower
values. The reason for this is that the evolution of the CV slows down
considerably for values of the donor mass of about $0.3\, M_\odot.$ Magnetic
breaking ceases, and the further evolution of the orbit is governed by
gravitational radiation, which is generally a slower process. Thus, if we could
establish that the donor had a mass of less than $0.2\, M_\odot,$ then the spin
down time would have to be larger than $\sim 10^9$~years.

\section{Discussion}

Nearby \iae\ remnants provide us the best opportunity to identify the progenitor
systems.  The discovery of an ex-companion star in these remnants would be
direct evidence of an SD progenitor.  Hence, any process that can hide such
companions is important. \citet{distefano11} discuss one such mechanism, the
spin-up/spin-down channel, in which the recently widowed companion has $10^5$ to
$10^9$ years to evolve during the spin-down phase and before the explosion.

Revisiting the available data on SNR 0509-67.5, we demonstrate that the current
magnitude limit of the {\em HST} observations, $V=26.9$ mag, is relatively
shallow to reject an SD progenitor that had to spin-down before an explosion.
However, deeper imaging observations with the {\em HST WFC3} can achieve
magnitude limits that are more meaningful for ex-companion star searches. Figure
1 demonstrates that an $M_V=11$ mag search, $V\approx29.5$ mag at the LMC's
distance, would identify or rule-out all C/O and He-core WDs for spin-down times
of $\leq10^7$ years and $M\leq 1 M_{\odot}$ WDs for $\leq10^8$ years. These
observations would also be sensitive to all ex-companion main-sequence stars
with $M\geq0.2 M_{\odot}$ for spin-down times of up to $10^9$ years. Even deeper
limits can be achieved with the {\em James Webb Space Telescope}.  If the donor
star is detected in such observations, this would be an important first, that
would allow us to model the progenitor and determine the spin-down time. If no
ex-companions are detected, this would suggest either a double-degenerate progenitor system
or a double detonation event with an undetectably faint present-day helium-core WD companion
\citep{nomoto80,woosley94,ruiter11,sim12}.
In either case, the result would be the most concrete clue to the
nature of any SN~Ia progenitor, and would provide the most direct insight yet to
the solution of the SNe~Ia progenitor puzzle.

Another complication in the studies of \iae\ remnants arises due to the uknown
center of the explosion. Based on hydrodynamic simulations, \citet{dohm96}
demonstrate that in the presence of a nonuniform external density region there
can be a significant shift in the apparent geometric center of a remnant away
from the explosion center. If the remnant is smaller than the width of the
density transition, it maintains a circular shape but the apparent center
shifts. \citet{borkowski06} and \citet{williams11} detect a large asymmetry in
the infrared brightness profile of SNR 0509-67.5; one hemisphere of this remnant
is brighter than the other side by a factor of five in the {\em Spitzer} 24
$\mu$m images. This brightness difference is most likely due to the forward
shock running into material that is several times denser than in other places
\citep{williams11}. Therefore, there is indirect evidence from the infrared
observations that the instellar medium in the explosion site is nonuniform and
the apparent center of this remnant has likely shifted from its original
location.

A comparison of the {\em HST} images of SNR 0509-67.5 \citep[see Figure 1
in][]{schaefer12} and the hydrodynamic simulations \citep[see Figures 10 and 11
in][]{dohm96} show that the apparent center is likely to shift toward the
low-density side. Hence, the real explosion site for SNR 0509-67.5 is likely to
the right of the current geometric center. \citet{schaefer12} consider such a
shift based on the geometric appearance and the ellipticity of the remnant.
However, \citet{dohm96} argue that the geometric center corrected for the
ellipticity of the remnant may not represent the actual center of a Ia remnant
and that detailed hydrodynamical simulations are required to constrain the
actual explosion site (S. Reynolds 2012, private communication).  There are
several targets, e.g. stars B, C, E, and I, to the right of the 3$\sigma$ error
circle as measured by \citet{schaefer12}.  The absolute magnitudes and the $V-I$
colors of these stars are consistent with 0.5-1 $M_{\odot}$ main-sequence stars.
Hence, one of these could be the ex-companion star to SNR 0509-67.5 that evolved
as a CV during the spin-down phase of the exploding WD. Until hydrodynamic
simulations for this remnant are available, a main-sequence ex-companion cannot
be ruled out for SNR 0509-67.5.

\section{Conclusions}

Recent studies based on the delay time distribution of \iae\ \citep{maoz10} and
the merger rate of double WD systems \citep{brown11,badenes12} suggest that
double WDs may contribute significantly to \iae\ events. Therefore, there is a
tendency in the community toward searching for more evidence for such progenitor
systems. One way to identify a double WD progenitor is to search for and
rule-out ex-companion stars in \iae\ remnants. Proving that something does not
exist is obviously difficult, and such claims require extra-ordinary evidence.
In this paper, we revisited the evidence for absence of an ex-companion star in
the young LMC \iae\ remnant SNR 0509-67.5. We demonstrate that in the
spin-up/spin-down model, the WD needs $10^5-10^9$ years before an Ia explosion,
which lets the companion to evolve into either a faint WD (if the companion
starts as a subgiant) or a low-mass main-sequence star (if the companion starts
as a main-sequence star). The current limits on the ex-companions are not
stringent enough to rule out faint WDs and $M<0.5 M_{\odot}$ main-sequence stars
in SNR 0509-67.5. Hence, the absence of evidence for an ex-companion star in
this remnant, as well as other \iae\ remnants, does not provide the evidence of
absence for such companions.

Future deep imaging observations can detect or rule out the majority of the C/O
WD ex-companions and low-mass main-sequence stars for plausiable spin-down times
of $\leq10^9$ years. If a companion is detected, this would be a first, and it
would constrain the uncertain spin-down times for WDs near the Chandrasekhar
mass. A non-detection can, in principle, rule out the SD channel even in the
spin-up/spin-down models. However, until such observations are obtained, we
consider an SD progenitor for SNR 0509-67.5 still possible.

\acknowledgements
We thank C. Badenes for useful discussions. We also thank the anonymous referee
for pointing out that double detonations from stably transferring (i.e.,
non-merging) He-rich degenerate donors are not ruled out.

\end{document}